%% file: main.tex
\documentclass[letterpaper,twocolumn,10pt]{article}
\usepackage{usenix2019_v3}

\usepackage{tikz}
\usepackage{amsmath}

\input{config}

\usepackage{filecontents}

\begin{document}

\clearpage
\pagenumbering{arabic}
\twocolumn
\date{}

\title{\arch: A Security Architecture with \textsc{Cu}stomizable and Resilient Enclaves}

\author{
\rm{Raad Bahmani, Ferdinand Brasser, Ghada Dessouky,}\\
\rm{Patrick Jauernig, Matthias Klimmek, Ahmad-Reza Sadeghi, Emmanuel Stapf}\\
Technische Universit\"at Darmstadt, Germany\\
\normalsize{\{raad.bahmani, ferdinand.brasser, ghada.dessouky, patrick.jauernig,}\}\\
\normalsize{\{matthias.klimmek, ahmad.sadeghi, emmanuel.stapf}\}\normalsize{@trust.tu-darmstadt.de}}

\maketitle

\input{sections/abstract}

\renewcommand{\paragraph}[1]{\noindent\textbf{#1}}

\input{sections/introduction}

\input{sections/adversary}

\input{sections/req_analysis}
\input{sections/design}

\input{sections/implementation}

\input{sections/sec_con}
\input{sections/evaluation}
\input{sections/related_work}

\input{sections/conclusion}

{\bibliographystyle{plain}
{\footnotesize
	\bibliography{bib}}}

\end{document}

%% file: config.tex
\usepackage{fancyhdr}
\usepackage{xspace}
\usepackage{cleveref}
\usepackage{makecell}
\usepackage{booktabs}
\usepackage{multirow}
\usepackage{enumitem}
\usepackage{calc}
\usepackage{caption}
\usepackage{array}
\IfFileExists{stix2.sty}{
	\usepackage[notext,not1]{stix2}
}

\microtypecontext{spacing=nonfrench}

\usepackage{tikz}

\newcommand{\arch}{\textsc{Cure}\xspace}

\newcommand{\master}{parent\xspace}
\newcommand{\masters}{parents\xspace}
\newcommand{\slave}{child\xspace}
\newcommand{\slaves}{childs\xspace}

\newcommand\blfootnote[1]{%
	\begingroup
	\renewcommand\thefootnote{}\footnote{#1}%
	\addtocounter{footnote}{-1}%
	\endgroup
}

%% file: sections/abstract.tex
\begin{abstract}
Security architectures providing Trusted Execution Environments (TEEs) have been an appealing research subject for a wide range of computer systems, from low-end embedded devices to powerful cloud servers. The goal of these architectures is to protect sensitive services in isolated execution contexts, called \textit{enclaves}.
Unfortunately, existing TEE solutions suffer from significant design shortcomings. 
First, they follow a \textit{one-size-fits-all} approach offering only a single enclave \textit{type}, however, different services need flexible enclaves that can adjust to their demands.
Second, they cannot efficiently support emerging applications (e.g., Machine Learning as a Service), which require secure channels to peripherals (e.g., accelerators), or the computational power of multiple cores. 
Third, their protection against cache side-channel attacks is either an afterthought or impractical, i.e., no fine-grained mapping between cache resources and individual enclaves is provided.

In this work, we propose \arch, the first security architecture, which tackles these design challenges by providing different types of enclaves: 
(i)~\emph{sub-space} enclaves provide vertical isolation at all execution privilege levels, (ii)~\emph{user-space} enclaves provide isolated execution to unprivileged applications, and (iii)~\emph{self-contained} enclaves allow isolated execution environments that span multiple privilege levels.
Moreover, \arch enables the exclusive assignment of system resources, e.g., peripherals, CPU cores, or cache resources to single enclaves.
\arch requires minimal hardware changes while significantly improving the state of the art of hardware-assisted security architectures. We implemented \arch on a RISC-V-based SoC and thoroughly evaluated our prototype in terms of hardware and performance overhead. \arch imposes a geometric mean performance overhead of 15.33\% on standard benchmarks.

\end{abstract}

%% file: sections/introduction.tex
\vspace{-0.2cm}
\section{Introduction}
\label{sec:introduction}

For decades, software attacks on modern computer systems have been a persisting challenge leading to a continuous arms race between attacks and defenses. The ongoing discovery of exploitable bugs in the large code bases of commodity operating systems have proven them unsuitable for reliable protection of sensitive services~\cite{GPZ18_1, GPZ18_2}. This motivated various hardware-assisted security architectures integrating \textit{hardware security primitives} tightly into the System-on-Chip (SoC).
Capability-based systems, such as CHERI~\cite{woodruff2014cheri}, CODOMs~\cite{vilanova2014codoms}, IMIX~\cite{frassetto2018imix}, or HDFI~\cite{song2016hdfi}, offer fine-grained protection through (in-process) sandboxing, however, they cannot protect against privileged software adversaries (e.g., a malicious OS). 
In contrast, security architectures providing Trusted Execution Environments (TEE) enable isolated containers, also called \textit{enclaves}. Enclaves allow for a coarse-grained but strong protection against adversaries in privileged software layers. TEE architectures have been proposed for a variety of computing platforms
\blfootnote{This paper will be published in the proceedings of the 30th USENIX Security Symposium (USENIX Security '21 ).}
\footnote{TEE architectures for resource-constrained embedded systems (e.g., Sancus~\cite{noorman2013sancus}, TyTAN~\cite{brasser2015tytan}, TrustLite~\cite{koeberl2014trustlite} or TIMBER-V~\cite{timberv}) are not the subject of this paper.}
, in particular for modern high-performance computer systems, e.g., industry solutions like Intel SGX~\cite{sgxref}, AMD SEV~\cite{amd_sev}, ARM TrustZone~\cite{arm-trustzone}, or academic solutions such as Sanctum~\cite{costan2016sanctum}, Sanctuary~\cite{sanctuary}, Keystone~\cite{keystone}, or Komodo~\cite{komodo} to name some.

In this paper, we focus on TEE architectures for modern high-performance computer systems. 
We investigate the shortcomings of existing TEE architectures and propose an enhanced and significantly more flexible TEE architecture with a prototype implementation for the open RISC-V architecture. \\

\noindent\textbf{Deficiencies of existing TEE architectures.}
So far, existing TEE architectures have adopted a \emph{one-size-fits-all} enclave approach. 
They provide only one \textit{type} of enclave requiring applications and services to be adapted to these enclaves' features and limitations, e.g., Intel SGX restricts system calls of its enclaves and thus, applications need to be modified when being ported to SGX which produces additional costs. Additional efforts like Microsoft's Haven framework~\cite{haven} or Graphene~\cite{graphene} are needed to deploy unmodified applications to SGX enclaves. 
Moreover, today, we are using diverse services that process sensitive data, e.g., payment, biometric authentication, smart contracts, speech processing, Machine Learning as a Service (MLaaS), and many more. Each service imposes a different set of requirements on the underlying TEE architecture. One important requirement concerns the ability to securely connect to devices. For example on mobile devices, privacy-sensitive data is constantly collected over various sensors, e.g., audio~\cite{voiceguard}, video~\cite{sonka2014image}, or biometric data~\cite{bio_auth}. On cloud servers, massive amounts of sensitive data are aggregated and used to train proprietary machine learning models, often outside of the CPU, offloaded to hardware accelerators~\cite{gpu_ml}.  
However, TEE architectures such as SGX~\cite{sgxref}, SEV~\cite{amd_sev} and Sanctum~\cite{costan2016sanctum}, do not consider secure I/O at all, solutions such as Keystone~\cite{keystone} would require additional hardware to support DMA-capable peripherals, solutions like Graviton~\cite{graviton} require hardware changes at the peripheral side. TrustZone~\cite{arm-trustzone}, Sanctuary~\cite{sanctuary} and Komodo~\cite{komodo} cannot bind peripherals directly to individual enclaves.

Another important requirement imposed on TEE architectures is an adequate and practical protection against side-channel attacks, e.g., cache~\cite{brasser2017software, cache_attack_btb} or controlled side-channel attacks~\cite{controlled_sc, severed_amd, van2017telling}. Current TEE architectures either do not include cache side-channel attacks in their threat model, like SGX~\cite{sgxref}, or TrustZone~\cite{arm-trustzone}, only provide impractical solutions which heavily influence the OS, like Sanctum~\cite{costan2016sanctum}, or do not consider controlled side-channel attacks, e.g., SEV~\cite{amd_sev}.  
We will elaborate on the related work and the problems of existing TEE architectures in detail in \Cref{sec:related_work}. 

\noindent\textbf{This work.} In this paper, we present a TEE architecture, coined \arch, that tackles the problems of existing solutions with a cost-effective and architecture-agnostic design. \arch offers multiple types of enclaves: 
(i)~sub-space enclaves that isolate only parts of an execution context, 
(ii)~user-space enclaves, which are tightly integrated into the operating system, and 
(iii)~self-sustained enclaves, which can span multiple CPU-cores and privilege levels. 
Thus, \arch is the first TEE architecture offering a high degree of freedom in adjusting enclave boundaries to fulfill the individual functionality and security requirements of modern sensitive services such as MLaaS. \arch can bind peripherals, with and without DMA support, exclusively to individual enclaves. Further, it provides side-channel protection via flexible and fine-grained cache resource allocation.  
\\\\
\noindent\textbf{Challenges.} Building a TEE architecture with the described properties comes with a number of challenges. 
(i)~New hardware security primitives must be developed that allow enclaves to adapt to different functionality and security requirements. 
(ii)~Even though the security primitives should allow flexible enclaves, they must not require invasive hardware modification, which would impede cross-platform adoption.
(iii)~While the changes in hardware should remain small, performance overhead for managing enclaves in software must be minimized.
(iv)~Protections against the emerging threat of microarchitectural attacks in form of side-channel and transient-execution attacks must be considered in the design for all types of enclaves. 
\noindent\textbf{Contributions.} Our design of \arch and its implementation on the RISC-V platform tackles all these challenges. To summarize, our main contributions are as follows:\\

\begin{itemize}[nolistsep]
	\item We present \arch, our novel architecture-agnostic design for a flexible TEE architecture which can protect unmodified sensitive services in multiple enclave types, ranging from enclaves in user space, over sub-space enclaves, to self-contained (multi-core) enclaves which include privileged software levels and support enclave-to-peripheral binding. 
	\item We introduce novel hardware security primitives for the CPU cores, system bus and shared cache, requiring minimal and non-invasive hardware modifications.
	\item We prototype \arch for the open RISC-V platform using the open-source Rocket Chip generator~\cite{asanovic2016rocket}.
	\item We evaluate \arch's hardware and software components in terms of added logic and lines of code, and \arch's performance overhead on an FPGA and cycle-accurate simulator setup using micro- and macrobenchmarks.
\end{itemize}

%% file: sections/adversary.tex
\vspace{-0.4cm}
\section{System Assumptions}
\label{subsec:platform}
\arch targets a modern high-performance multi-core system, with common performance optimizations like data and instruction caches, a Translation Lookaside Buffer (TLB), shared caches, branch predictors, respective instructions to flush the core-exclusive resources, and a central system bus that connects the CPU with the main memory (over a dedicated memory controller) and various peripherals.

\noindent\textbf{System bus and peripherals.} 
The system bus connects the CPU to a plethora of system peripherals over a fixed set of hardwired peripheral controllers. The peripherals range from storage, communication, and input devices to specialized compute units, e.g., hardware accelerators~\cite{google_tpu}.
The CPU interacts with peripherals using parts of the internal peripheral memory which are mapped to the address space of the CPU, called Memory-Mapped I/O (MMIO). We assume that the CPU can nullify the internal memory of a peripheral to sanitize its state. Every access from the CPU to a peripheral is decoded in the system bus and delegated to the corresponding peripheral. The CPU acts as a \textit{\master} on the system bus, whereas the peripherals (and main memory) act as \textit{\slaves} that respond to requests from a \master. However, MMIO is not sufficient for some peripherals where large amounts of data need to be shared with the CPU since the CPU needs to copy the data from the main memory to the peripheral memory. Therefore, these peripherals are often connected to the system bus as \textit{\masters} over Direct Memory Access (DMA) controllers, allowing them to directly access the main memory.
To cope with resource contention in these complex interconnects, system buses also incorporate arbitration mechanisms to schedule the establishment of \master-\slave connections when multiple bus requests occur simultaneously.

\begin{figure}
	\centering
	\includegraphics[width=0.9\linewidth]{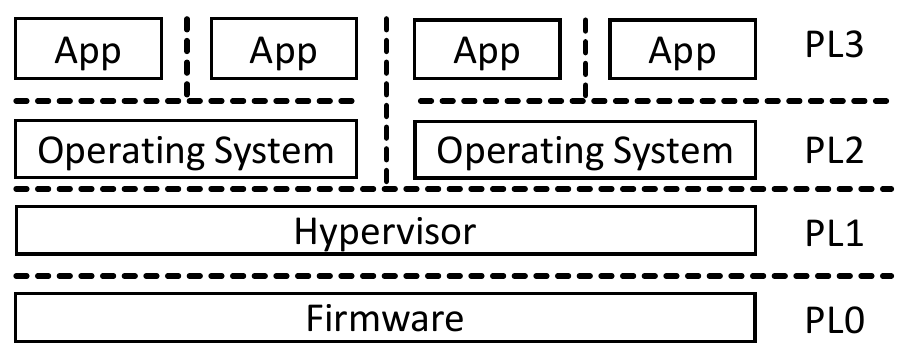}
	\caption{Software privilege levels (PL): user space, kernel space \& dedicated levels for hypervisor \& firmware.}
	\label{fig:priv_level}
\end{figure}

\noindent\textbf{Software privilege levels.} We assume the CPU supports the privilege levels (PLs) as shown in~\Cref{fig:priv_level}. In line with modern processors (Intel~\cite{intel_manual}, AMD~\cite{amd_manual} or ARM~\cite{arm_manual}), we assume a separation between a user-space layer (PL3) and a more privileged kernel-space layer (PL2), which is performed by the MMU (configured by PL2 software) through virtual address spaces. The CPU may support a distinct layer for hypervisor software (PL1) to run virtualized OS in Virtual Machines (VMs), where the separation to PL2 is performed by a second level of hardware-assisted address translation~\cite{virt_hype}. Lastly, we assume a highly-privileged layer (PL0) which contains firmware that performs specific tasks, e.g., hardware emulation or power management.

We assume that the system performs secure boot on reset, whereas the first bootloader stored in CPU Ready-Only Memory (ROM), verifies the firmware through a chain of trust~\cite{tbb_arm}. After verification, the firmware starts execution from a predefined address in the firmware code and loads the current firmware state from non-volatile memory (NVM) where it is stored encrypted, integrity- and rollback-protected. The cryptographic keys to decrypt and verify the firmware state are passed by the bootloader which loads the firmware into Random-access Memory (RAM). Rollback protection can be achieved, e.g., by making use of non-volatile memory with Replay Protected Memory Block (RPMB) partitions or by using eFuses as secure monotonic counters~\cite{rollback_arm}. When a system shutdown is performed, the firmware stores its state in the NVM, encrypted and integrity- and rollback-protected.

\vspace{-0.3cm}
\section{Adversary Model}
\label{subsec:adv_model}
Our adversary model adheres to the one commonly assumed for TEE architectures, i.e., a strong software-only adversary that can compromise all software components, including the OS, except a small software/microcode Trusted Computing Base (TCB) which configures the hardware security primitives of the system, manages the enclaves and which is inherently trusted~\cite{sgxref, arm-trustzone, komodo, sanctuary, keystone, costan2016sanctum}. 

We assume that the goal of the adversary is to leak secret information from the TCB or from a victim enclave. An adversary with full control of the system software can inject own code into the kernel (PL2) and even into the hypervisor (PL1). This allows the adversary, with full access to the TCB interface used for setting up enclaves, to spawn malicious processes and even enclaves. Even though the adversary cannot change the firmware code (which uses secure boot), memory corruption vulnerabilities might still be present in the code and be exploitable by the adversary~\cite{davidson2013fie}. In addition, we assume that an adversary is able to compromise peripherals from software to perform DMA attacks~\cite{sang2011attacks, markettos2019thunderclap}.

We assume the underlying hardware to be correct and trusted, and hence, exclude attacks that exploit hardware flaws~\cite{rowhammer-paper, tang2017clkscrew}. We also do not assume physical access, and thus, fault injection attacks~\cite{biehl2000differential}, physical side-channel attacks~\cite{koch_time, mangard2008power} or the physical connection of malicious peripherals are out of scope. We do not consider Denial-of-Service (DoS) attacks in which the adversary starves an enclave since an adversary with control over the OS can shut down the complete system trivially. As standard for TEE architectures, \arch does not protect from software-exploitable vulnerabilities in the enclave code but prevents their exploitation from compromising the complete system.

%% file: sections/req_analysis.tex
\vspace{-0.3cm}
\section{Requirements Analysis}
\label{sec:requirements}
To provide customizable, practical and strongly-isolated enclaves, \arch must fulfill a number of security and functionality requirements. We list them in the following section, and show in ~\Cref{sec:sec_con} how \arch fulfills the security requirements. In~\Cref{sec:implementation} and~\Cref{sec:evaluation}, we demonstrate how the functionality requirements are met.

\subsection{Security Requirements (SR)}
\label{subsec:s_req}
\begin{enumerate}[nolistsep, label=\textbf{SR.\arabic*:}, ref={SR.\arabic*}, leftmargin=0ex, labelsep=\widthof{~}, itemindent=\widthof{SR.1:~\hspace{.125ex}}] 
	\item \label{req:srenclaveprot1}\textbf{Enclave protection.} Enclave code must be integrity-protected when at rest, and inaccessible for an adversary when executed. All sensitive enclave data must remain confidential and integrity-protected at all times. An enclave must be protected from adversaries on all software layers (PL3-PL0), other potentially malicious enclaves, and DMA attacks~\cite{sang2011attacks, markettos2019thunderclap}. 
	\item \label{req:srhwprimitive2}\textbf{Hardware security primitives.} The protection of the enclaves must be enforced by secure hardware components which can only be configured by the software TCB.
	\item \label{req:srmintcb3}\textbf{Minimal software TCB.} The TCB must be protected from adversaries in all software layers (PL3-PL0) and minimal in size to be formally verifiable, i.e., a few KLOCs~\cite{sel4_formal}.
	\item \label{req:srsidechannel4}\textbf{Side-channel attack resilience.} Mitigations against the most relevant software side-channel attacks must be available, namely, side-channel attacks on cache resources~\cite{primeprobe, flushreload, cache_attack_btb, cache_attack_tlb}, controlled side-channel attacks~\cite{controlled_sc, severed_amd, van2017telling} and transient-execution attacks~\cite{Kocher2018spectre, van2018foreshadow, chen2018sgxpectre, spectre_ng, spectre_fallout, spectre_zombie, spectre_ridl, van2020lvi}. 
\end{enumerate}

\subsection{Functionality Requirements (FR)}
\label{subsec:f_req}
\begin{enumerate}[nolistsep, label=\textbf{FR.\arabic*:}, ref={FR.\arabic*}, leftmargin=0ex, labelsep=\widthof{~}, itemindent=\widthof{FR.1:~\hspace{.125ex}}]
	\item \label{req:frdynenclave1}\textbf{Dynamic enclave boundaries.} The trust boundaries of an enclave must be freely configurable such that enclaves at different privilege levels can be supported.
	\item \label{req:frperiphbinding3}\textbf{Enclave-to-peripheral binding.} Secure communication between enclaves and selected system peripherals, e.g., when offloading sensitive machine learning tasks to hardware accelerators~\cite{gpu_ml}, must be explicitly supported.
	\item \label{req:frsmallhw2}\textbf{Minimal hardware changes.} The hardware changes required to integrate the proposed security primitives into a commodity SoC (cf.~\Cref{subsec:platform}) must be minimal, no invasive changes to CPU internals must be required to enable a higher adoption of \arch in future platforms.
	\item \label{req:frperfoverhead4}\textbf{Reasonable performance overhead.} The performance overhead incurred during enclave setup and run time must be minimized and must not render the computer system impractical for certain uses cases or degrade user experience.
	\item \label{req:frconfprotmech5}\textbf{Configurable protection mechanisms.} Protection mechanisms against cache side-channel attacks must be applicable dynamically at run time and on a per-enclave basis.
\end{enumerate}

%% file: sections/design.tex
\vspace{-0.3cm}
\section{Design of the \arch Architecture}
\label{sec:design}
\begin{figure}[h]
	\centering
	\includegraphics[width=\linewidth]{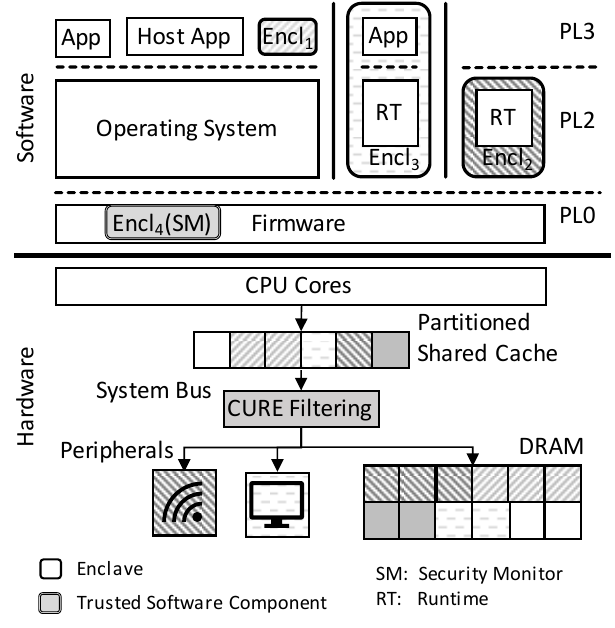}
	\caption{\arch privilege levels and enclave types, namely, user-space enclaves (Encl\textsubscript{1}), kernel-space enclaves (Encl\textsubscript{2}, Encl\textsubscript{3}) and sub-space enclaves (Encl\textsubscript{4}).}
	\label{fig:design_sw}
\end{figure}

\arch provides a novel design that addresses the requirements described above and provides a TEE architecture with strongly-isolated and highly customizable enclaves, which can be adapted to the requirements of the services they protect. 
Unlike other TEE architectures, which only provide a single enclave-type, \arch allows to freely define enclave boundaries and thus, different enclaves can be constructed, as shown in~\Cref{fig:design_sw}.
First, in~\Cref{subsec:design_setup}, we describe the ecosystem around \arch. Then, we elaborate on the different enclave types in~\Cref{subsec:design_sw}. 
\arch's key component enabling this flexible enclave construction is its enclave ID-based access control in the system bus which manages all per-enclave resource mappings, e.g, peripherals or main memory, indicated by the different background patterns in~\Cref{fig:design_sw} and~\Cref{fig:design_hw}. 
Our hardware primitives are presented in~\Cref{subsec:design_hw}.

\vspace{-0.3cm}
\subsection{\arch Ecosystem}
\label{subsec:design_setup}
The ecosystem around \arch consists of device vendors which produce the devices implementing \arch, device users and service providers. Some services contain sensitive data (from the users and/or the service provider) and thus, must be protected. In \arch, sensitive services are either split into a sensitive and a non-sensitive part, which get included into an enclave and an user-space app (called host app), respectively, or alternatively, integrated entirely into an enclave, requiring only minimal modifications at the service. In the later case, the host app is only needed to trigger the enclave. Initially, the enclave binary does not contain sensitive data.

For every enclave, the service provider creates a configuration file which contains the enclave's requirements regarding system resources (e.g., memory, caches or peripherals), a version number and an enclave label $L_{encl}$. Enclave binary, configuration file and host app are bundled and deployed by the service provider over an app store (e.g., Google Play Store) which is operated by a third party (e.g., Google). The label $L_{encl}$ is globally unique in the app store.

Every service provider creates an asymmetric key pair $SK_p$ and $PK_p$, and a public key certificate $Cert_p$, which is signed by the app store operator. Using the secret key $SK_p$, the service provider signs the enclave binary and configuration file ($Sig_{encl}$) and attaches it, together with $Cert_p$, to the app bundle. $Cert_p$ can later be used on the device to verify $Sig_{encl}$. For this, a certificate chain $Chain_p$ up to the root certificate of the app store operator must be present on the device. When the service provider wants to update an enclave, a new signature must be created and the version number in the configuration file updated which prevents rollbacks to older (possibly flawed) versions of an enclave~\cite{tee_rollback_attack}.

A device vendor creates a unique asymmetric key pair $SK_d$ and $PK_d$ for each device, which is provisioned to the device during production, and a public key certificate $Cert_d$ signed by the device vendor which can later be used to prove the legitimacy of the device in a remote attestation scheme. For this, the service provider must obtain a certificate chain $Chain_d$ up to the root certificate of the device vendor. When a device was compromised, $Cert_d$ can also be revoked.

\vspace{-0.3cm}
\subsection{Customizable and Resilient Enclaves}
\label{subsec:design_sw}
\arch supports enclaves that protect user-space processes (Encl\textsubscript{1}), run in the kernel space (Encl\textsubscript{2}) or span the kernel and user space (Encl\textsubscript{3}). However, an enclave does not necessarily include all code of a privilege level, e.g., an enclave can only comprise parts of the firmware code (Encl\textsubscript{4}).

\subsubsection{Enclave Management}
\label{subsubsec:enc_mgmt}
Before describing the different enclave types supported by \arch, we give an overview on \arch's enclave management.

\noindent\textbf{Security monitor.} All \arch enclaves are managed by the software TCB, called \textit{Security Monitor (SM)}, as in other TEE architectures~\cite{costan2016sanctum,keystone}. As indicated in~\Cref{fig:design_sw}, the SM itself represents an enclave which is part of the firmware. As described in~\Cref{subsec:platform}, we assume a system that performs a secure boot on reset, verifies the firmware (including the SM) and then jumps to the entry point of the SM. Further, we assume that the SM has already loaded its rollback protected state $S_{sm}$ into the volatile main memory. The SM state contains $SK_d$, $PK_d$, $Cert_d$, $Chain_p$ and a structure $D_{encl}$ for each enclave installed on the device.

\noindent\textbf{Enclave installation.}
When an enclave is deployed to the device, the SM first verifies the signature $Sig_{encl}$ using $Cert_p$ and $Chain_p$. Then, the SM creates a new enclave meta-data structure $D_{encl}$ and stores $L_{encl}$, $Sig_{encl}$ and $Cert_p$ in it. Moreover, the SM creates an enclave state structure $S_{encl}$ which is used to persistently store all sensitive enclave data. The SM also creates an authenticated encryption key $K_{encl}$ which is used to protect the enclave state when it is stored to disk or flash memory. $K_{encl}$ and $S_{encl}$ are also stored in $D_{encl}$. Initially, $S_{encl}$ only contains an authenticated encryption key $K_{com}$ created by the SM, which is used by the enclave to encrypt and integrity protect data communicated to the untrusted OS, and a monotonic counter. The enclave meta-data structure $D_{encl}$ also contains a monotonic counter used to rollback protect the enclave state.

\noindent\textbf{Enclave setup \& teardown.}
The setup of an enclave is always triggered by the corresponding host app. After the OS loads the enclave binary and configuration file, it performs a context switch to the SM. The SM identifies the enclave by the label $L_{encl}$ and begins the enclave setup by (1)~configuring the hardware security primitives (\Cref{subsec:design_hw}) such that one or multiple continuous physical memory regions (according to the configuration file) are exclusively assigned to the enclave in order to isolate the enclave from the rest of the system software. Since the binary and configuration file are loaded from untrusted software, their integrity must always be verified using $Sig_{encl}$ and $Cert_p$. Assigning physical memory regions is inevitable when providing enclaves which are able to execute privileged software (kernel-space enclave), since this allows the enclave to control the MMU. Thus, virtual memory cannot be used to effectively isolate the enclave. (2)~After enclave verification, the SM configures the hardware primitives to assign also the rest of the system resources, e.g., cache or peripherals, to the enclave according to the configuration file. All assigned resources are also noted in $D_{encl}$. Moreover, the SM assigns an identifier to the enclave which is stored in $D_{encl}$ and which is unique for every enclave currently active on the device. The SM can manage up to $N$ (implementation defined) enclaves in parallel. We provide more details on the meaning of the enclave identifier in~\Cref{subsec:design_hw}. (3)~In the last step, the enclave state $S_{encl}$ is restored, i.e., loaded from disk or flash memory,  decrypted and verified using $K_{encl}$, and then copied to the enclave memory such that it is accessible during enclave runtime. The SM also checks that the monotonic counter in $S_{encl}$ matches the counter stored in $D_{encl}$.
 
The SM configures all interrupts to be routed to the SM while an enclave is running. Thus, the SM fully controls the context switches into and out of an enclave. While the SM is executed, all interrupts on the CPU core executing the SM are disabled. All other cores remain interrupt responsive.
In \arch, hardware-assisted hyperthreading is disabled during enclave execution to prevent data leakage through resources shared between the hardware threads. Alternatively, all hardware threads of a CPU core could also be assigned to the enclave if the enclave code benefits from parallelization. In the reminder of the paper, we assume that hyperthreading is disabled during enclave runtime.

After the setup is complete, the SM jumps to the entry point of the enclave. During the enclave teardown, which can be triggered by the host app or the enclave itself, the SM securely stores the enclave state (using $K_{encl}$), while incrementing the monotonic counters in $S_{encl}$ and $D_{encl}$, removes all enclave data from the memory and caches and reconfigures the hardware primitives.

\noindent\textbf{Enclave execution.} At run time, enclaves can access services provided by the SM over its API, e.g., to dynamically increase the enclave's memory or to receive an integrity report which the SM creates by signing $Sig_{encl}$ with $SK_d$ and by attaching $Cert_d$. The integrity report is then send to the service provider by the enclave. Subsequently, using $Chain_d$, the service provider can perform a remote attestation of the enclave. Only if the attestation succeeds, the service provider provisions sensitive data to the enclave. More complex remote attestation schemes~\cite{Intel-Remote_Attestation} could also be implemented. 

Enclaves might use services of the untrusted OS which do not require access to the plain sensitive enclave data, e.g., file or network I/O. For those cases, an enclave can utilize $K_{com}$, which is part of $S_{encl}$, to protect its sensitive data. \arch also allows multiple enclaves to share encrypted sensitive data over the OS. However, the required key exchange is assumed to be performed over the back ends of the service providers and thus, out-of-scope for \arch.

Every enclave which includes a cryptographic library can also create own keys (apart from $K_{com}$) and store them in $S_{encl}$. Thus, enclaves can also implement key rotation, revocation or recovery schemes which is, however, the responsibility of the service provider and thus, out-of-scope for \arch.

On every enclave setup/teardown and context switch in and out of an enclave, the SM flushes all core-exclusive cache resources, i.e., the data cache, the TLB and the BTB, thereby preventing information leakage across execution contexts.

\subsubsection{User-space Enclaves}
\label{subsubsec:user_enclave}
User-space enclaves (Encl\textsubscript{1} in ~\Cref{fig:design_sw}) comprise a complete user-space process. 

\noindent\textbf{OS integration.} The key characteristic of a user-space enclave is its tight integration into the OS, i.e., it relies on the OS for memory management, exception/interrupt handling and other services provided through syscalls (e.g., file system or network I/O). The OS schedules user-space enclaves like normal user-spaces processes, only that the context switches in and out of the enclave are intercepted by the SM. The OS's services are used by all user-space enclaves which prevents code duplication. Moreover, user-space enclaves do not contain management software, leading to smaller binaries. 

\noindent\textbf{Controlled side-channel defenses.} 
In controlled side-channel attacks, the adversary gains information about an enclave's execution state by observing usage of resources managed by the OS, predominantly page tables\cite{controlled_sc, severed_amd, van2017telling}. 
\arch defends against these attacks by moving the page tables of user-space enclaves into the enclave memory. 
More subtle controlled side-channel attacks exploit the fact that the enclave's interrupt handling is performed by the OS~\cite{van2018nemesis}. \arch also mitigates these attacks by allowing each enclave to register trap handlers to observe its own interrupt behavior, and act accordingly if a suspicious behavior is detected~\cite{tsgx, deja_vu}. 

\noindent\textbf{Limitations \& usage scenarios.} A user-space enclave cannot run higher-privileged code, e.g., device drivers. Thus, all sensitive data shared with a peripheral has to be processed by drivers in the untrusted OS and thus, is unprotected if not encrypted. Hence, user-space enclaves are unable to protect sensitive services which interact with devices like sensors or GPUs. Instead, user-space enclave are beneficial when protecting short-living services that can rely on encrypted data transmission, e.g., One Time Password (OTP) generators, payment services, digital key services and many more.

\subsubsection{Kernel-space Enclaves}
\label{subsubsec:design_ks_enc}
Kernel-space enclaves can comprise only the kernel space (Encl\textsubscript{2}), or the kernel and user space (Encl\textsubscript{3}).

\noindent\textbf{Providing OS services.} The key characteristic of a kernel-space enclave is its capability to run code bare-metal on a CPU core in the privileged (PL2) software layer or even in the hypervisor level (PL1) if available. Thus, OS services, e.g. memory management, can be implemented inside the enclave in a runtime (RT) component (\Cref{fig:design_sw}). This results in less resource sharing with the untrusted OS, and thus, it is easier to protect against controlled side-channel attacks~\cite{controlled_sc, van2017telling, van2018nemesis}. Moreover, by including device drivers into the RT, a secure communication channel to peripherals can be established. Furthermore, kernel-space enclaves provide more computational power since \arch allows to run kernel-space enclaves across multiple cores. In \arch, peripherals can either be assigned exclusively to a single enclave, by the SM, at enclave setup or shared between different enclaves and/or the OS. The peripheral's internal memory is flushed by the SM when (re-)assigned to a new entity to prevent information leakage ~\cite{pietro2016cuda, lee2014stealing, zhou2017vulnerable}.  

\noindent\textbf{Protecting virtual machines.} \arch's ability to include the kernel space into the enclave allows the construction of enclaves that encapsulate complete virtual machines (VMs). VMs are not self-contained but rely on memory and peripheral management services provided by a hypervisor, which makes the VM enclave vulnerable to controlled side-channel attacks~\cite{amd_io, amd_sev}. 
\arch mitigates this by moving the VM page tables into the enclave memory and including unmodified complete drivers into the enclave to avoid dependencies on the untrusted hypervisor~\cite{xen_hype, kvm_hype}. As for other kernel-space enclaves, peripherals are temporarily assigned to VM enclaves by the SM. Again, before a peripheral is reassigned, its internal memory is sanitized by the SM. 

\noindent\textbf{Limitations \& usage scenarios.} Sensitive services can be ported to kernel-space enclaves without changing them. However, in contrast to user-space enclaves, an enclave RT needs to be added which increases the binary size, adds development overhead and increases the memory consumption. Moreover, the CPU cores selected for the enclave first have to be freed from pending processes, detached from the OS and the RT booted on them. 
Nevertheless, kernel-space enclaves are required when protecting services which heavily rely on peripheral communication, e.g., authentication services using biometric sensors, ML services collecting input data over sensors or offloading computations to accelerators, DRM services or in general services which require secure I/O.

\subsubsection{Sub-space Enclaves}
\label{subsubsec:design_subspace_encl}
In \arch, enclave trust boundaries can be freely defined which allows to construct fine-grained enclaves that only include parts of the software residing in a privilege level, therefore called sub-space enclaves.

\noindent\textbf{Shrinking the TCB.} Sub-space enclaves are especially appealing when constructed in the highest privilege level (PL0) of the system (Encl\textsubscript{4} in~\Cref{fig:design_sw}). In \arch, sub-space enclaves are used to isolate the SM from the firmware code to protect against exploitable memory corruption vulnerabilities that might be present in the firmware code~\cite{davidson2013fie}. Moreover, hardware countermeasures, described in~\Cref{subsec:design_hw}, are used to prevent the firmware code from accessing the SM data or hardware primitives. Ultimately, this minimizes the software TCB in \arch, as opposed to other TEE architectures that rely on a software TCB containing all code in the highest privilege level, i.e., EL3 (ARM) or the machine level (RISC-V), e.g., TrustZone~\cite{arm-trustzone}, Sanctuary~\cite{sanctuary}, Sanctum~\cite{costan2016sanctum}, Keystone~\cite{keystone}. 

\subsection{Hardware Security Primitives}
\label{subsec:design_hw}
\begin{figure}
	\centering
	\includegraphics[width=0.8\linewidth]{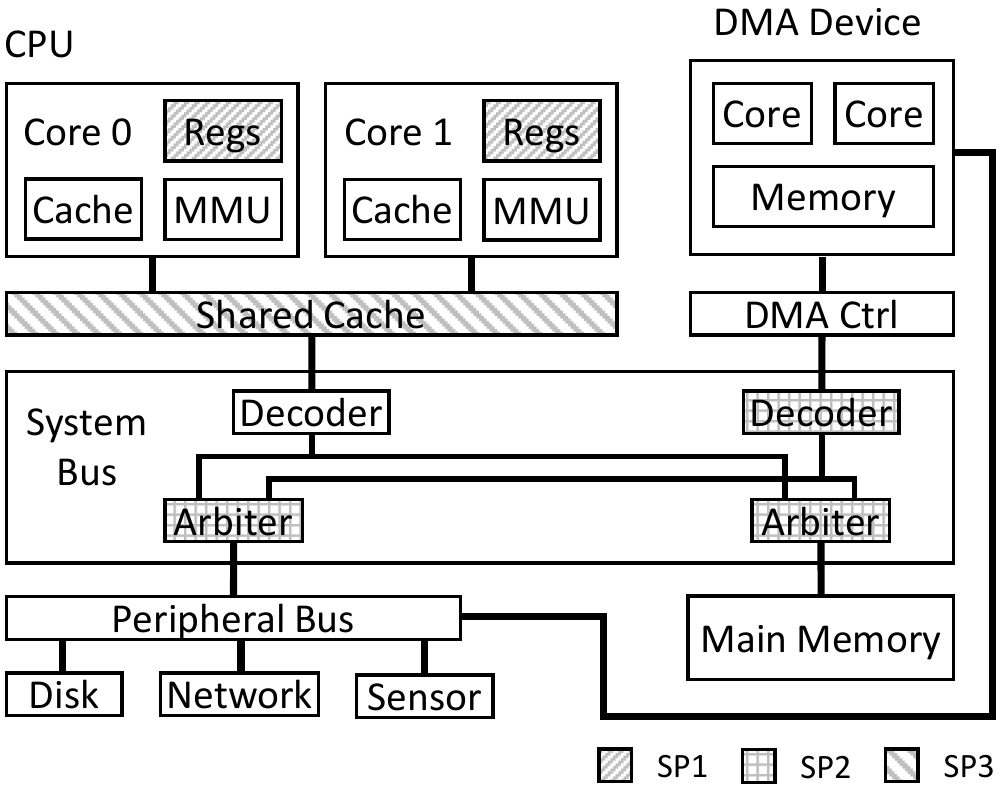}
	\caption{\arch Security Primitives (SPs), added at core register files (SP1), system bus (SP2) and shared cache (SP3).}
	\label{fig:design_hw}
\end{figure}

To provide \arch's customizable enclaves, new security primitives (SP) are needed in hardware. Our SPs augment the register file of each CPU core (SP1), the system bus (SP2) and the shared cache (SP3).~\Cref{fig:design_hw} shows where \arch's SPs integrate in a modern system as assumed in~\Cref{subsec:platform}.

\vspace{-0.3cm}
\subsubsection{Defining Enclave Execution Contexts (SP1)}
\label{subsubsec:design_hw_sp1}
\noindent\textbf{Enclave ID register.} In \arch, enclave execution contexts are defined using IDs, which are saved in a register that is added to every CPU core of the system (SP1). At any point in time, the value of this register, called \texttt{eid} (enclave ID) register, indicates which enclave a core currently executes. The \texttt{eid} registers are set by the SM during enclave setup, teardown and any context switch in and out of an enclave, thus, enabling flexible configuration of enclave boundaries.

Whenever an enclave is set up, the SM assigns it an unused ID. In contrast to the constant enclave labels $L_{encl}$ (\Cref{subsubsec:enc_mgmt}) , which are globally unique, an enclave ID is only valid as long as the enclave is loaded in memory. When an enclave is torn down, the ID gets freed and can be assigned to the next enclave. Constant IDs are only assigned to the SM and all untrusted software. The number of different IDs ($N$) that can be stored in \texttt{eid} defines how many enclaves can run in parallel (\Cref{subsubsec:enc_mgmt}). However, the total number of enclaves that can be deployed is not restricted.

\noindent\textbf{Propagating the enclave ID.} The enclave ID is propagated through the entire system and used in the SPs to perform access control on the system resources. We incorporate the enclave ID in the bus protocol between the CPU, shared cache and system bus. In protocols like AMBA AXI4/ACE~\cite{amba_axi_ace}, the de facto on-chip communication standard, no protocol extensions are required since the bus channels provide optional user-defined signals which can be utilized to transmit the enclave ID in bus transactions. In our \arch prototype, we extend the TileLink protocol~\cite{tilelink} by an enclave ID signal, which we describe in more detail in~\Cref{sec:implementation}.

\subsubsection{Access Control on the Bus (SP2)}
\label{subsubsec:design_hw_sp2}
In order to isolate enclaves and assign peripherals to them, access control mechanisms need to be implemented in hardware. As described in~\Cref{subsec:platform}, the system bus represents the central gateway of a computer system that connects bus \masters (CPU or DMA devices) with bus \slaves (peripherals or the main memory) and routes all their transactions. \arch leverages this centralization and further extends it to perform access control on \master-\slave transactions (SP2 in~\Cref{fig:design_hw}). Incorporating carefully crafted access control at the system bus, with latency and performance in mind, reduces the overall hardware costs significantly.

\noindent\textbf{Enclave memory isolation.} One key task of a TEE architecture is enforcing strong isolation of the enclave code and data in the main memory. In \arch, this is achieved by performing access control in the arbiter logic in front of the main memory chip, as shown in~\Cref{fig:design_hw}. This requires adding new registers and control logic to the already existing arbiter, which can only be configured (over MMIO) by the SM to assign memory regions to enclaves. Whenever the CPU requests access to a memory address, the arbiter uses the enclave ID signal, which is sent within the bus transaction, to verify if the enclave currently executing is allowed to access the memory region. At access violation, the memory access is prevented and an interrupt is triggered by the system bus, which is handled by the SM. 
Incorporating the required logic for this access control at the main memory side, instead of the CPU side, reduces the additional registers and logic required, which would otherwise be duplicated for every CPU core, as we show in~\Cref{subsec:system-eval}.

\noindent\textbf{Assigning peripherals to enclaves.} The CPU interacts with peripherals over peripheral memory mapped to the CPU address space (MMIO). In \arch, access control on the MMIO memory is performed using registers and control logic added to the arbiter at the peripheral bus. The SM assigns the MMIO region of every peripheral either to one enclave exclusively or to multiple enclaves/OS by configuring the arbiter registers. Access control is then performed in the added hardware logic based on the enclave ID signal of a bus transaction. Incorporating this logic at the CPU side would have increased the hardware costs because of per-core duplication.

\noindent\textbf{DMA protection.} Peripherals which share large amounts of data with the CPU typically access the main memory directly over a DMA controller. \arch must protect enclaves from DMA attacks~\cite{sang2011attacks, markettos2019thunderclap} and also allow to assign DMA-capable peripherals to enclaves. To achieve this, \arch adds registers and control logic to the decoder in front of every DMA device. These registers define which memory regions the DMA device is allowed to access. Whenever a DMA device gets assigned to an enclave, the SM updates the device registers accordingly. 
Adding the required logic at the \slave arbiters would increase the hardware costs because enclave IDs would also need to be assigned to the DMA devices which would result in additional logic for ID comparison.

By assigning dedicated memory regions to an enclave and a DMA-capable peripheral, and by assigning the MMIO memory regions of that peripheral exclusively to the enclave, \arch achieves an enclave-to-peripheral binding. Since neither the OS nor any other enclave can access the memory regions over which the bound enclave and peripheral communicate, no encryption or authentication schemes are required.

\subsubsection{On-Demand Cache Partitioning (SP3)} 
\label{subsubsec:cache-design}
\arch's enclave management (in~\Cref{subsubsec:enc_mgmt}) mitigates side-channel attacks on core-exclusive resources, such as the L1 cache, by flushing all such structures at every enclave context switch. Nevertheless, this still leaves enclaves vulnerable to cross-core attacks on the shared last-level cache~\cite{irazoqui2015s, flushreload, kayaalp2016high}. 
However, vulnerability to these sophisticated attacks depends on whether the enclave code performs memory accesses dependent on sensitive data. While algorithms and implementations can be constructed leakage-resilient~\cite{ohrimenko2016oblivious,ahmad2018obliviate}, this is not directly applicable to any given application code, and thus, we provide on-demand per-enclave cache partitioning in \arch.

Security guarantees for cache side-channel resilience can be provided in hardware by either enforcing strict partitioning of resources across the different enclaves~\cite{Liu16,Wang16,kiriansky2018dawg} or deploying randomization-based cache schemes~\cite{Newcache16,Liu14}. 
Nevertheless, these schemes either reduce the cache resources available for an enclave or incur additional access latency. This results in an inevitable performance overhead on the protected as well as unprotected software. The additional security guarantee, along with its resulting performance cost, is not usually required for all enclaves and largely depends on the use case.

Thus, \arch addresses these diverse enclave requirements and incorporates on-demand way-based partitioning of the shared cache (SP3 in~\Cref{fig:design_hw}). This allows that cache partitioning is enabled and configured individually and dynamically for each enclave at setup and runtime. Each cache way can be allocated exclusively to an enclave. 
Access control on the enclave ID signal of the memory access transaction is used to permit the enclave to access (read/write or even evict) a cache way, thus ensuring strict isolation. However, when this cache isolation is not enabled for an enclave, only read/write access control on the owner enclave of each cache line is performed. This defends against a privileged adversary that can access cached enclave memory by mapping it into its own address space. As each cache line is owned by a single enclave at any point in time, access control on cache lines corresponding to shared memory between enclaves and the OS is a challenge. To address this, the SM flushes relevant cache lines at context switches between an enclave and the OS while managing shared-memory communication.

We deploy way-based partitioning because it is the least extensive in terms of hardware modifications. However, \arch provides the necessary infrastructure and mechanisms (by identifying each enclave and propagating this throughout the system bus and shared cache) to incorporate more sophisticated side-channel-resilient cache designs~\cite{ceaser, scattercache, hybcache}.

%% file: sections/implementation.tex
\section{Prototyping \arch on \textsc{RISC-V}}
\label{sec:implementation}
While \arch is architecture-agnostic and can be ported to other ISAs, we prototype it here for a RISC-V system based on the open-source Rocket Chip generator~\cite{asanovic2016rocket}. We describe next our \arch instantiation, followed by details on the implemented enclave types and hardware security primitives. 

\begin{figure}
	\centering
	\includegraphics[scale=1.0]{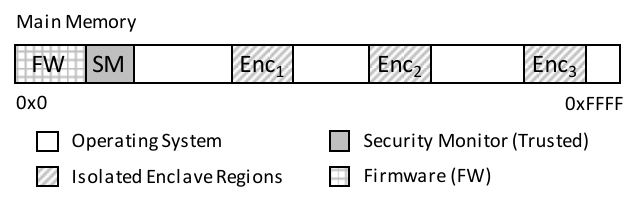}
	\caption{Physical memory layout of our \arch prototype.}
	\label{fig:mem_layout}
\end{figure}

\noindent\textbf{RISC-V System-on-Chip platform.}
We build a RISC-V System-on-Chip (SoC) using the Rocket Chip generator~\cite{asanovic2016rocket}. For prototyping, we equipped the SoC with multiple in-order Rocket cores, in line with prototyping efforts in related work \cite{costan2016sanctum}. Each Rocket core has one hart (representing a hardware thread), an own MMU, BTB, TLB and L1 cache. 
The SoC also contains a system bus which connects the cores to system peripherals (over the peripheral bus) and system main memory. We integrate a shared L2 cache~\cite{sifive_l2} between the system bus and the main memory. A DMA device is connected to the system bus as a bus \master. As a result, this SoC resembles our assumed platform shown in~\Cref{fig:design_hw}, except that the L2 cache is integrated as a last-level cache after the system bus.

We implement our prototype on this SoC aiming to maintain minimum hardware and no additional latency. 
We use 4 bits to represent the enclave ID, i.e., our prototype can distinguish 16 ($N$) enclaves, where ID \texttt{0} is statically assigned to the OS, ID \texttt{0xF} to the Security Monitor (SM) and ID \texttt{0xE} to the firmware (explained in~\Cref{subsubsec:crossbar}). The remaining 13 IDs can be freely assigned to enclaves. We assign one continuous physical memory region to each enclave, resulting in the memory layout shown in~\Cref{fig:mem_layout}. We choose to assign only one region per enclave to simplify our prototype and minimize the induced hardware overhead. The CURE design, however, also allows for multiple continuous regions per enclave.
The SM and firmware memory regions are adjacent since they are both deployed as part of the bootloader~\cite{riscv_pk}. All regions not assigned to an enclave, SM or the firmware, belong to the OS. Supporting more enclaves in parallel is possible if the additional hardware overhead is acceptable.

\noindent\textbf{Software stack.} The Rocket core supports three software privilege levels (user, supervisor and machine). Hypervisor support is still a work-in-progress~\cite{riscv_priv_spec} and thus, we do not consider it in our prototype. In the supervisor level, we use an OS consisting of a modified Linux LTS kernel 4.19 with a Busybox 1.29.3 environment. We add a custom kernel module which performs security-uncritical tasks during the enclave setup. We implement the SM in the machine level as a sub-space enclave to separate it from the firmware which runs in the same privilege level.

\begin{figure}[h]
	\centering
	\includegraphics[scale=1.0]{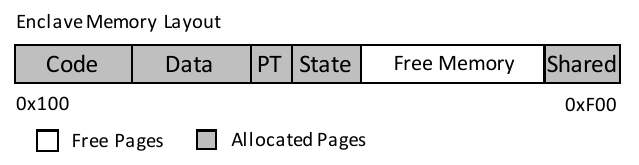}
	\caption{\arch enclave memory layout consisting of the code \& data pages, page tables (PT), the enclave state (State) and the shared memory (Shared).}
	\label{fig:enc_mem_layout}
\end{figure}

\noindent\textbf{Cryptographic underpinnings.} 
In the implemented \arch prototype, we use Ed25519~\cite{ed25519} as the digital signature scheme for the signing and verification of the enclave signature $Sig_{encl}$ and the integrity report used for remote attestation, as described in~\Cref{subsubsec:enc_mgmt}. Thus, $SK_d$/$PK_d$ and $SK_p$/$PK_p$ are Ed25519 key pairs. The public key certificates $Cert_d$ and $Cert_p$ are implemented in the X.509 format. In our \arch prototype, the certificate chains $Chain_d$ and $Chain_p$ required to verify $Cert_d$ and $Cert_p$ are, for the sake of simplicity, represented by two Ed25519 public keys. As described in~\Cref{subsubsec:enc_mgmt}, $Chain_p$ is included in the SM, whereas $Chain_d$ is required at the service provider. The enclave state $S_{encl}$ and enclave data communicated with the OS are protected through authenticated encryption, using the keys $K_{encl}$ and $K_{com}$, respectively. We use AES-GCM from libtomcrypt 1.18.2.~\cite{libtomcrypt} as the authenticated encryption scheme and include it in the SM. Moreover, we also add it to our implemented enclaves, such that the enclaves can create additional keys. Consistent with~\Cref{subsubsec:enc_mgmt}, the SM holds a meta-data structure $D_{encl}$ for each enclave which contains $Cert_p$, $Sig_{encl}$, $K_{encl}$ and $S_{encl}$, whereas $K_{com}$ is part of $S_{encl}$.

\subsection{Software \arch Enclaves}
\label{subsec:impl_sw_enclave}
Our \arch prototype implements user-space enclaves, kernel-space enclaves and sub-space enclaves and thus, fulfills requirement \texttt{\ref{req:frdynenclave1}} (\Cref{subsec:f_req}). In the following, we describe the enclave memory layout and give implementation details on each enclave type.

\subsubsection{Enclave Memory Layout}
In our prototype, each enclave is assigned a continuous physical memory region which is allocated during enclave setup using Linux's Contiguous Memory Allocator (CMA). The enclave memory layout is shown in~\Cref{fig:enc_mem_layout}. At the lowest address, the enclave code and data pages are loaded by the OS.
The enclave page tables are only stored in the enclave memory while the memory management is performed by the untrusted OS. During the enclave setup, the SM loads the enclave state $S_{encl}$ into the enclave memory. The free memory space is used for dynamic memory allocation. 
The memory region at the highest address is used for the communication between enclave and OS. Since our prototype allows one continuous memory region per enclave, the shared memory region is either assigned to the communicating enclave or to no enclave, which automatically assigns the region to the OS. When the enclave is set up, the address of the shared memory region is communicated to the OS via the return value of the SM call. The enclave is informed by storing the address information on the stack of the enclave. The size of the enclave state and shared region can be freely set, we set them to 64 bytes and 4 KB, respectively.

\subsubsection{Security Monitor}
\label{subsubsec:impl_sec_mon}
We implement the SM as a sub-space enclave (Enc\textsubscript{5} in~\Cref{fig:design_sw}) separated from the firmware in memory (\Cref{fig:mem_layout}), which is enforced by the hardware security primitives. However, this leaves the firmware with access to the security-critical machine level registers \texttt{eid}, which we added, and \texttt{mtvec}, which holds the base address of the trap vector that the core jumps to after an interrupt.
To prevent the firmware from configuring these registers, we implement a hardware mechanism that ensures that the \texttt{eid} and \texttt{mtvec} registers can only be written to when the \texttt{eid} register is set to the SM ID (\texttt{0xF}). The \texttt{eid} register is, in turn, set to \texttt{0xF} by the hardware when performing a context switch to machine mode that traps in the SM. 

\subsubsection{User-space Enclaves}
\label{subsubsec:impl_user_encl}
\noindent\textbf{Memory management.} Since the memory management of the user-space enclave (Enc\textsubscript{1} in~\Cref{fig:design_sw}) is performed by the untrusted OS, we include the enclave page tables in the enclave memory, to prevent page table based attacks~\cite{controlled_sc, severed_amd, van2017telling}. During enclave setup, the OS creates the page tables exactly as for a normal process. However, the OS turns off demand paging and maps all code and data pages to prevent page faults during enclave execution. The page tables are then handed to the SM which verifies their validity.
Moreover, the SM verifies that the supverisor address translation and protection (\texttt{satp}) register, which holds the address of the root page table, points into the enclave memory.
Subsequently, the page tables are copied to the enclave memory. Once the enclave is setup, the OS cannot alter the page tables anymore. When the dynamic allocation of memory leads to a page fault, the OS creates a new page table entry and passes it to the SM which includes it into the page tables.

\noindent\textbf{Syscalls.} Our prototype provides enclaves which can use OS services, e.g., file or network I/O, over Linux syscalls which trap in the SM. We include AES-GCM into the enclaves to encrypt and integrity-protect sensitive data shared with the OS, using $K_{com}$.
Enclaves are always exited through the SM which is enforced by clearing the machine exception delegation (\texttt{medeleg}), machine interrupt delegation (\texttt{mideleg}), supervisor exception delegation (\texttt{sedeleg}) and supervisor interrupt delegation (\texttt{sideleg}) registers during enclave setup. During run time, the enclave can register custom trap handlers which are called by the SM before switching to the OS after an interrupt. Thus, the enclave can observe its own interrupt behavior and detect suspicious behavior caused by interrupt-based side-channel attacks~\cite{deja_vu, van2018nemesis}.

\subsubsection{Kernel-space Enclaves}
\label{subsubsec:impl_kernel_encl}
Our \arch prototype supports kernel-space enclaves with and without user space (Enc\textsubscript{3} and Enc\textsubscript{2} in~\Cref{fig:design_sw}). We use an Linux LTS kernel 4.19, which currently on RISC-V does not support a suspension mode, as the enclave RT.
 
\noindent\textbf{Allocating resources.} When an enclave is set up, the custom kernel module unmounts the driver modules of all peripherals requested by the enclave. Then, the SM performs the security-critical tasks of the enclave setup, as described in~\Cref{subsubsec:design_ks_enc}. When the enclave binary is successfully verified, the kernel module shuts down the core(s) reserved for the enclave using the Linux hotplugging mechanism. Next, a switch to the SM is performed which jumps to the entry point of the enclave RT in order to boot the RT on all reserved cores. At enclave shutdown, the SM performs the cleanup, and all freed cores are reintegrated into the OS. Then, the kernel module remounts the driver modules.

\noindent\textbf{Enclave-OS communication.}
Since our \arch prototype allows one memory region per enclave, access to a shared region needs to be requested at the SM which then assigns the shared region to the requesting party (sender). Once the sender is finished accessing the shared region, the SM assigns the shared region to the receiver and notifies the receiver about new data in the shared region using an inter-processor interrupt.
In contrast to the user-space enclave, only external interrupts are trapped in the SM during kernel-space enclave execution which is enforced by configuring the \texttt{medeleg} and \texttt{sedeleg} registers during the enclave setup. All interrupts triggered by the enclave cores are handled by the RT. 

\begin{figure}[t!]
	\centering
	\includegraphics[width=\linewidth]{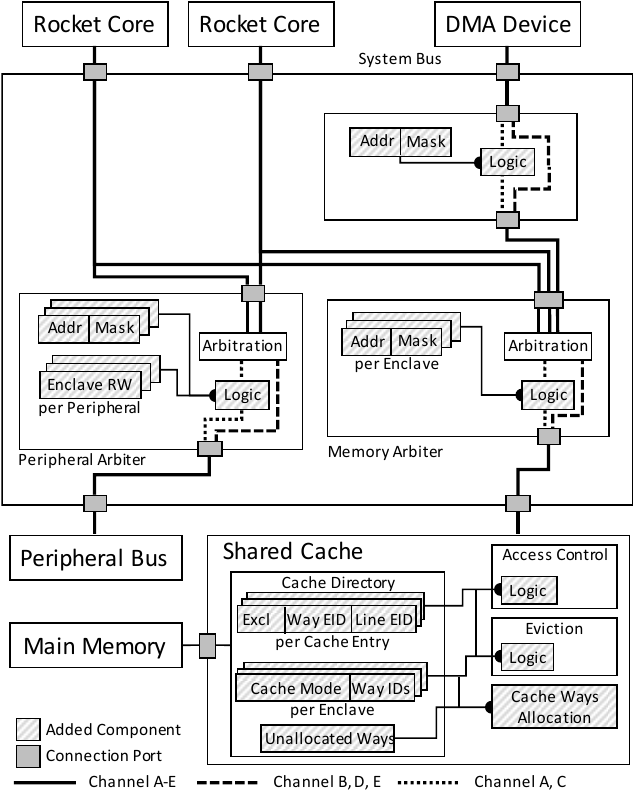}
	\caption{\arch prototype implementation using Rocket Chip.}
	\label{fig:impl}
\end{figure}

\subsection{Hardware Security Primitives}
\label{subsec:impl_hw}
We describe next, how we modify the Rocket Chip to implement \arch's hardware security primitives (\Cref{subsec:design_hw}).

\subsubsection{Extending the TileLink Protocol}
\label{subsubsec:tilelink}
We modify the Rocket core such that on every memory access, the \texttt{eid} register value is sent as part of the issued bus transaction. This also includes transactions issued by the PTW (page table walker) during the page table walk when performing address translations. Thus, if a malicious enclave modified its own page tables to point to a memory region outside of the enclave memory, the PTW transactions are blocked by the access control mechanisms on the system bus.

TileLink~\cite{tilelink} is the default bus protocol used on the Rocket Chip to connect on-chip components. TileLink specifies five channels (\texttt{A} - \texttt{E}). When connecting a \master to the system bus which contains an internal cache, all five channels are needed to implement the TileLink coherence protocol (TL-C). When a \master does not require cache coherency, only the \texttt{A} and \texttt{D} channels are needed (TL-UL/UH). In our RISC-V SoC, the Rocket cores and the DMA devices are connected over TL-C since they contain L1 caches. 

We extend the TileLink protocol by a 4-bit \texttt{eid} signal to propagate the enclave ID. The \texttt{eid} signal is only added to the \texttt{A} and \texttt{C} channels which transport the memory read and write transactions from the \masters (CPU and DMA devices) to the system bus and \slaves (peripherals and main memory), respectively. All other channels remain unmodified. 

\subsubsection{System Bus Access Control}
\label{subsubsec:crossbar}
We implement \arch's access control mechanisms in the system bus by adding registers and control logic at the memory and peripheral arbiters and the ports connecting DMA devices. The hardware changes are shown in~\Cref{fig:impl}, exemplary for a system containing two cores, one DMA device and multiple peripherals. All newly added components are connected to the control bus of the system and thus, are configurable by the SM over MMIO. We omit the control bus in~\Cref{fig:impl} for the sake of clarity. Our implementation supports enclave-to-peripheral binding and thus, fulfills \texttt{\ref{req:frperfoverhead4}}. Moreover, in contrast to related work~\cite{cotret2012bus,coburn2005seca}, all access control is performed in parallel to arbitration, thus, guaranteeing execution in a single clock cycle without incurring additional latency.

\noindent\textbf{Performing access control.} The added registers hold memory ranges defined by a 32-bit base address (\texttt{Addr}) and a 32-bit mask (\texttt{Mask}), and are used by the control logic to perform access control on every memory transaction using the \texttt{eid} and \texttt{address} signals.
Access control is only performed on channels with a \master-to-\slave direction (channels \texttt{A} and \texttt{C}). At access violation, the transaction is redirected (with all-zero data) to an unused, zero-initialized memory region. Thus, all forbidden transactions write/read zeros to/from the unused memory region.  
An adversary enclave might fill L1 with malicious data which could get flushed with SM privileges during enclave context switch. To prevent this, we modify the core such that on every switch to the SM, the L1 is flushed before the \texttt{eid} register is set. We connect the system bus to the peripheral and interrupt bus. This allows the SM to configure the added registers and control logic, and trigger an interrupt upon access violation which is handled by the SM.

\noindent\textbf{Memory arbiter.} We add 15 registers to the memory arbiter, one for each enclave (13), the SM and the firmware. Each register defines the memory region assigned to each execution context. For the enclaves, the control logic verifies that transactions only target the assigned region. For the SM, no access control is performed. The OS is allowed to access all regions except the ones specified in registers of the arbiter. The firmware is allowed to access its own and the OS regions which is why a static ID needs to be assigned to the firmware.

\noindent\textbf{Peripheral arbiter.} We add two registers per peripheral to the arbiter of the peripheral bus. One covers the MMIO region of the peripheral, and the other 32-bit register contains a bitmap that defines read and write permissions for every enclave.

\noindent\textbf{DMA port.} We add a register at every port which connects a DMA device. In \arch, a DMA device is exclusively assigned to a single enclave at one point in time. In our prototype, a DMA device accesses the main memory but not other peripherals. If specific use cases, e.g. PCI peer-to-peer transactions~\cite{gpudirect}, must be supported, additional registers need to be added to specify multiple allowed memory regions. Together with the peripheral arbiter, this fulfills \ref{req:frperiphbinding3}.

\subsubsection{L2 Cache Partitioning}
\label{subsubsec:cache}
For cache side-channel resilience, we implement way-based flexible cache partitioning for the shared L2 (last-level) cache~\cite{sifive_l2} in our prototype. We leverage the \texttt{eid}-extended TileLink memory transactions to detect when an enclave issues a cache request.

\textbf{Configurable partitioning.} We implement two modes of partitioning to allow enclaves to individually enable cache side-channel resilience. The first mode \texttt{CP-BASIC} performs rudimentary access control where each enclave is only permitted to access (hit) its own cache lines, but is free to evict cache lines from other ways. The second mode \texttt{CP-STRICT} provides more stringent security guarantees by allocating \textit{exclusively} one or more ways (across all cache sets) to the pertinent enclave. Only these cache ways can be accessed by the enclave to store or evict cache lines. This provides strict isolation between the cache resources of the different enclaves, thus, effectively blocking cache side-channel leakage, but reduces the cache resources available for the enclave. Depending on the enclave service requirements, the partitioning mode can be configured by the SM independently for each enclave at setup and during the enclave lifetime, thus, fulfilling~\texttt{\ref{req:frconfprotmech5}}.

\textbf{Access control.} We extend each cache entry metadata with a 4-bit \texttt{line-eid} register encoding the owner enclave of the cache line, as shown in in~\Cref{fig:impl}. We extend the cache lookup logic to generate a hit only when both tag as well as \texttt{eid} match for \texttt{CP-BASIC}, as opposed to usual tag matching. 

To support \texttt{CP-STRICT}, the cache ways directory is also extended with a 1-bit register \texttt{excl} that identifies whether each way is owned exclusively by an enclave, as well as a 4-bit \texttt{eid} register that identifies the owner enclave. The cache controller logic is augmented with a register-based lookup table that is indexed by the \texttt{eid}. It encodes with a single \texttt{mode} bit whether the corresponding enclave has \texttt{CP-STRICT} enabled and its allocated cache way indices.
In \texttt{CP-STRICT}, cache hits are only allowed in these cache ways.

\textbf{Eviction and replacement.} The L2 cache we use implements a pseudo-random replacement policy where any way is selected pseudo-randomly for eviction. We modify this to only select a way from the subset of ways allowed for each enclave. For enclaves with \texttt{CP-STRICT}, only ways exclusively allocated to it are used. For enclaves with \texttt{CP-BASIC}, all ways (except ways allocated exclusively to other enclaves) are used.

\textbf{Per-enclave cache allocation.} Unallocated way indices are maintained in a register vector. If an enclave with \texttt{CP-STRICT} enabled requests to exclusively own cache ways, the required ways are allocated if available and below the allowed maximum per enclave.

An inherent drawback of this partitioning technique is how the limited number of cache ways directly constrains the number of simultaneous enclaves that can have \texttt{CP-STRICT} enabled. However, this is only an implementation decision for our particular prototype, where more sophisticated cache designs~\cite{ceaser, scattercache, hybcache} can be integrated into \arch.

%% file: sections/sec_con.tex
\section{Security Considerations}
\label{sec:sec_con}
To protect from a strong software adversary, our instantiation of \arch must fulfill the security requirements introduced in~\Cref{subsec:s_req}. In the following section, we discuss how our prototype meets the requirements \texttt{\ref{req:srenclaveprot1}}, \texttt{\ref{req:srhwprimitive2}}, and \texttt{\ref{req:srsidechannel4}}, whereas we show the fulfillment of \texttt{\ref{req:srmintcb3}} in~\Cref{sec:evaluation}.

\subsection{Hardware Security Primitives (\texttt{SR.2})}
\label{subsec:sec_con_hw}
The enclave protection is enforced by hardware SPs at the system bus and L2 cache which are configured over MMIO. After the system is powered on and on every switch to the machine level, the CPU jumps to the trap vector whose address is stored in the \texttt{mtvec} register. The trap vector is included into the SM such that the boot process and context switches are overlooked by the SM. The \texttt{mtvec} register is assigned to the SM by coupling the access permission to the SM enclave ID (stored in the \texttt{eid} register) which is also assigned to the SM. The \texttt{eid} register is set by hardware during the context switch into the machine level. During boot, the SM assigns the SP MMIO regions exclusively to its own enclave ID.

\subsection{Enclave Protection (\texttt{SR.1})}
\label{sec:sec_con_sr1}
At rest, the enclave binaries are stored unencrypted in memory. However, during enclave setup, the SM verifies the binaries using digital signatures. Moreover, the L1 is flushed during setup/teardown to remove malicious or sensitive data from the cache. The communication between enclaves and the OS is controlled by the SM, so is the delegation of the shared memory address. Hardware-assisted hyperthreading is disabled during enclave execution. The enclave state, which is loaded during the setup process, is persistently stored by the SM using authenticated encryption, either in RAM as part of the SM state or evicted to flash/disk, and additionally rollback protected. During teardown, the SM removes all enclave data from the memory.

The SPs in hardware perform access control on physical addresses at the system bus. Thus, \arch protects from adversaries in privileged software levels (PL2 - PL0) and from off-core adversaries, e.g. peripherals performing DMA. The enclave data cached in the L1 during run time is protected by flushing it on all context switches. Data in the L2 cache is protected by assigning cache lines exclusively to enclaves. Since no enclave (except the SM), has elevated rights on the system, \arch also protects from malicious enclaves.

\subsection{Side-channel Attack Resilience (\texttt{SR.4})}
\label{subsec:side_channel}
\noindent\textbf{Cache side-channel attacks.} Side-channel attacks which target data in core-exclusive cache resources, i.e., in the L1~\cite{brasser2017software}, the BTB~\cite{cache_attack_btb} or the TLB~\cite{cache_attack_tlb}, are prevented by the SM by flushing the resources on all context switches. Side-channel attacks targeting data in the shared L2 cache~\cite{irazoqui2015s, flushreload, kayaalp2016high} are prevented through strict way-based cache partitioning.

\noindent\textbf{Controlled side-channel attacks.} Side-channel attacks on user-space enclaves which target page tables~\cite{controlled_sc, severed_amd, van2017telling} are prevented by including the page tables into the enclave memory and by mapping all enclave code and data pages before execution. The SM verifies the page tables and the base address of the root page table stored in the \texttt{satp} register. The hardware SPs prevent the page table walker (PTW) from performing forbidden memory access during the page table walk. Side-channel attacks exploiting interrupts~\cite{van2018nemesis} can be mitigated using trap handlers (\Cref{subsubsec:user_enclave}).

\arch provides cryptographic primitives in the user-space enclaves to encrypt and integrity-protect data shared with the OS. However, using OS services over syscalls always comprises a remaining risk of leaking meta data information~\cite{schuster2017beauty, ahmad2018obliviate} or of receiving malicious return values from the OS~\cite{checkoway2013iago}. In user-space enclaves, these attacks must be mitigated on the application level inside the enclave, e.g., by using data-oblivious algorithms~\cite{ahmad2018obliviate, ohrimenko2016oblivious} or by verifying the return values~\cite{checkoway2013iago}.
None of these attacks pose a threat to kernel-space enclave since all resources are handled by the enclave RT. However, on VM enclaves, the second level page tables need to be protected, as with user-space enclaves. Interrupt-based attacks can again be mitigated with custom trap handlers. No additional countermeasures are needed to protect the SM since the SM does not use a virtual address space or OS services and handles its own interrupts.

\noindent\textbf{Transient execution attacks.} The discovered transient execution attacks either mistrain the branch predictor~\cite{Kocher2018spectre, spectre_ng, chen2018sgxpectre}, rely on information leakage~\cite{van2018foreshadow} or malicious injections~\cite{van2020lvi} on the L1 cache, or rely on resources shared when using hardware-assisted hyperthreading~\cite{spectre_zombie, spectre_ridl, spectre_fallout, van2020lvi, sgaxe}. By disabling hyperthreading during enclave execution (or alternatively assigning all threads to the enclave) and flushing core-exclusive caches, \arch protects enclaves against the known transient execution attacks.

%% file: sections/evaluation.tex
\vspace{-0.4cm}
\section{Evaluation}
\label{sec:evaluation}
In the following section, we systematically evaluate our \arch prototype. First, we quantify the software and hardware modifications required to implement \arch. Next, we evaluate the performance of \arch's enclaves using microbenchmarks, and the overall performance overhead of \arch using generic RISC-V benchmark suites.

\vspace{-0.4cm}
\subsection{System Modifications}
\label{subsec:system-eval}
\begin{table}[h]
	\centering
	\renewcommand\theadalign{lr}
	\renewcommand\theadfont{\bfseries}
	\renewcommand\theadgape{\Gape[0pt]}
	\renewcommand\cellgape{\Gape[4pt]}
	\setlength{\abovecaptionskip}{5pt plus 0pt minus 0pt}
	\setlength{\belowcaptionskip}{0pt plus 0pt minus 0pt}
	\small
	\begin{tabular}{l|r}
		\thead{Component} & \thead{LOC} \\
		\hline
		Linux Kernel & 375 (modified) \\
		Custom Kernel Module & 200 \\
		Security Monitor & 544  \\
		SM Crypto-Library & 2586\\
		\hline
	\end{tabular}
	\caption{Lines of code required to implement \arch. SM Crypto-Library refers to the crypto library (part of tomcrypt) included in the Security Monitor.}
	\label{tab:loc}
\end{table}

\paragraph{Software changes and TCB.} 
Our implementation of \arch on RISC-V comprises of a slightly modified Linux LTS kernel 4.19, a custom kernel module, and our software TCB (SM). In \Cref{tab:loc}, the lines of code (LOC) are shown for each of the components, which indicate that the software changes required to implement \arch are minimal. Moreover, the SM only consists of around 3KLOC of code, whereas most (82.62\%) of the SM code consists of cryptographic primitives. Because of its minimal size, formal verification of the SM is possible~\cite{sel4_formal}, thus, fulfilling~\ref{req:srmintcb3}. Note that since \arch isolates the SM in an own sub-space enclave, \arch can achieve a smaller TCB size than other RISC-V security architectures~\cite{costan2016sanctum,keystone,timberv} which include all code in the machine level, i.e., the firmware code, in the TCB. In our implementation, the firmware code consists of 3286 LOCs. Thus, by isolating the SM in a sub-space enclave, we managed to cut the software TCB in half, where the actual management code is even less~(15.56\%).

Protecting a sensitive service in a user-space enclave requires to add a small custom library (10KB) to the service binary. For the kernel-space enclaves, management code (the enclave RT) must be added in addition. In our prototype, we use the Linux LTS kernel 4.19 as the RT which increases the size
of the service binary by 3MB. Custom RTs can further decrease this kernel-space enclave overhead. However, kernel-space enclaves will always have an increased binary size and memory consumption compared to user-space enclaves.

\noindent\textbf{Hardware overhead.} 
We evaluate the hardware overhead of our changes by synthesizing the generated Verilog descriptions using Xilinx Vivado tools targeting a Virtex UltraScale FPGA device.  
\Cref{tab:hw-overhead} shows a breakdown of the individual area overhead of the different modifications required to implement \arch. Overhead is represented in look-up tables (LUTs), the fundamental programmable logic blocks of FPGA devices, and registers.

\begin{table}[h!]
	\small
	\setlength{\abovecaptionskip}{5pt plus 0pt minus 0pt}
	\setlength{\belowcaptionskip}{-16pt plus 0pt minus 0pt}
	\begin{center}
		\begin{tabular}{l|l|l}
			\multirow{2}{*}{\textbf{Configuration}} & \textbf{LUTs} & \textbf{Registers}\\
			& \textbf{Overhead (+\%)} &  \textbf{Overhead (+\%)}\\
			\hline
			Baseline & $61,097$  & $28,012$ \\
			TileLink extension & $+211$ $(0.4\%)$ &  $+110$ $(0.4\%)$ \\
			\hline
			\hline
			\multicolumn{3}{c}{Access control extensions}\\
			\hline
			\hline
			Main memory & $+5,276$ $(8.6\%)$ & $+1,055$ $(3.8\%)$\\
			1 MMIO peripheral & $+248$ $(0.4\%)$ & $+107$ $(0.4\%)$\\
			1 DMA device & $+112$ $(0.2\%)$ & $+72$ $(0.3\%)$\\
			\hline
			\hline
			\multicolumn{3}{c}{On-demand cache partitioning}\\
			\hline
			\hline
			w/ L2 cache (baseline) & $+30,232$ & $+11,549$ \\
			w/ L2 cache partitioned & $+516$ ($1.7\%^\ast$) & $+214$ ($1.8\%^\ast$) \\
			\hline
		\end{tabular}
		\caption{Hardware overhead breakdown in LUTs and registers. Baseline setup consists of 2 Rocket cores without L2 cache. $^\ast$Overhead relative to the L2 cache (baseline).}
		\label{tab:hw-overhead}
	\end{center}
\end{table}

We compare in~\Cref{tab:hw-overhead} with a baseline configuration of 2 in-order Rocket cores (each with L1 cache). Extending the TileLink protocol throughout the system bus incurs a minimal overhead of $105$ LUTs per core relative to the baseline (211 LUTs for 2 cores). This overhead includes propagating the \texttt{eid} in tandem with memory access transactions through the MMU of every core, and is thus replicated for every additional core in the system.

In contrast, the rest of our modifications for performing access control at the system bus, including enclave-to-peripheral binding, are independent of the number of cores. Incorporating logic to perform access control for every MMIO peripheral utilizes an additional $248$ LUTs, and $112$ LUTs per DMA device. Each represent below $0.5\%$ overhead relative to a dual-core baseline SoC. Integrating an L2 cache into our baseline setup utilizes an additional $30,232$ LUTs. Applying our on-demand way-based partitioning to this cache costs only $516$ LUTs and $214$ registers, which is $1.8\%$ overhead relative to the L2 cache logic utilization itself, and $0.5\%$ relative to the entire SoC. Our area overhead evaluation results demonstrate that the hardware modifications required to achieve our fine-grained and customized enclave protection in \arch indeed incur minimal area overhead on both single- and multi-core architectures, thus fulfilling~\ref{req:frsmallhw2}.

\vspace{-0.4cm}
\subsection{Performance Evaluation}\label{subsec:perf}
We evaluate the performance of \arch using our FPGA-based setup coupled with cycle-accurate simulators. We conduct our experiments using micro and macro benchmarks for user-space and kernel-space enclaves, and compare them to unmodified user-space processes. We conduct 10 runs for each of the experiments.

\begin{table}[t]
	\centering
	\setlength{\abovecaptionskip}{5pt plus 0pt minus 0pt}
	\renewcommand\theadgape{\Gape[0pt]}
	\resizebox{\columnwidth}{!}{		
		\begin{tabular}{lrcrr}
			\textbf{Measurement} &       & \multicolumn{1}{c}{\textbf{\thead{Normal\\Process}}} & \textbf{\thead{User-Space\\Enclave}} & \textbf{\thead{Kernel-Space\\Enclave}} \\\hline
			\textbf{Setup:} &       & \multicolumn{1}{c}{0.741} & 23.918 & 413.726			 \\
			\multicolumn{2}{l}{\hspace{7mm}Binary Verification} & \multicolumn{1}{c}{-} & 21.824 & 218.975			 \\
			\multicolumn{2}{l}{\hspace{7mm}Others} & \multicolumn{1}{c}{0.741} & 2.094 & 194.750 \\
			\hline
			\textbf{Teardown:} &       & \multicolumn{1}{c}{0.065} & 23.531 & 103.517 \\
			\multicolumn{2}{l}{\hspace{7mm}Memory Cleaning} & \multicolumn{1}{c}{-} & 9.384 & 50.206 \\
			\multicolumn{2}{l}{\hspace{7mm}Others} & \multicolumn{1}{c}{0.065} & 14.147 & 53.311\\
			\midrule
			Context switch to OS &       & \multicolumn{1}{c}{0.008} & 0.025			 & 53.308 \\
			Context switch from OS &       & \multicolumn{1}{c}{0.078} & 0.084 & 194.747 \\
			Dynamic memory allocation &       & \multicolumn{1}{c}{0.003} & 0.020 & 0.005 \\
			OS communication &       & - & 0.020 & 0.049 \\\hline
		\end{tabular}
	}
	\caption{\arch performance overhead compared to a normal process on microbenchmarks in milliseconds.}
	\label{tab:micro}
\end{table}

\vspace{-0.4cm}
\subsubsection{Microbenchmarks}
\label{subsubsec:microbenchmarks}
For microbenchmarks (\Cref{tab:micro}), we measured important key aspects individually: setting up and tearing down an enclave, context switching with the OS, dynamic memory allocation, and communication via shared memory. 
We implement an application which performs the required tasks (without any additional logic) and run it as a normal Linux process, a user-space enclave and a kernel-space enclave (single core). The enclave setup is triggered by a host app in Linux which is the only purpose of the app. The enclave binary sizes therefore mainly correspond to the overhead produced by the enclave types, i.e., 10KB for the user-space enclave and around 3MB for the kernel-space enclave.

\begin{figure}[t!]
	\centering
	\setlength{\abovecaptionskip}{5pt plus 0pt minus 0pt}
	\includegraphics[width=\linewidth]{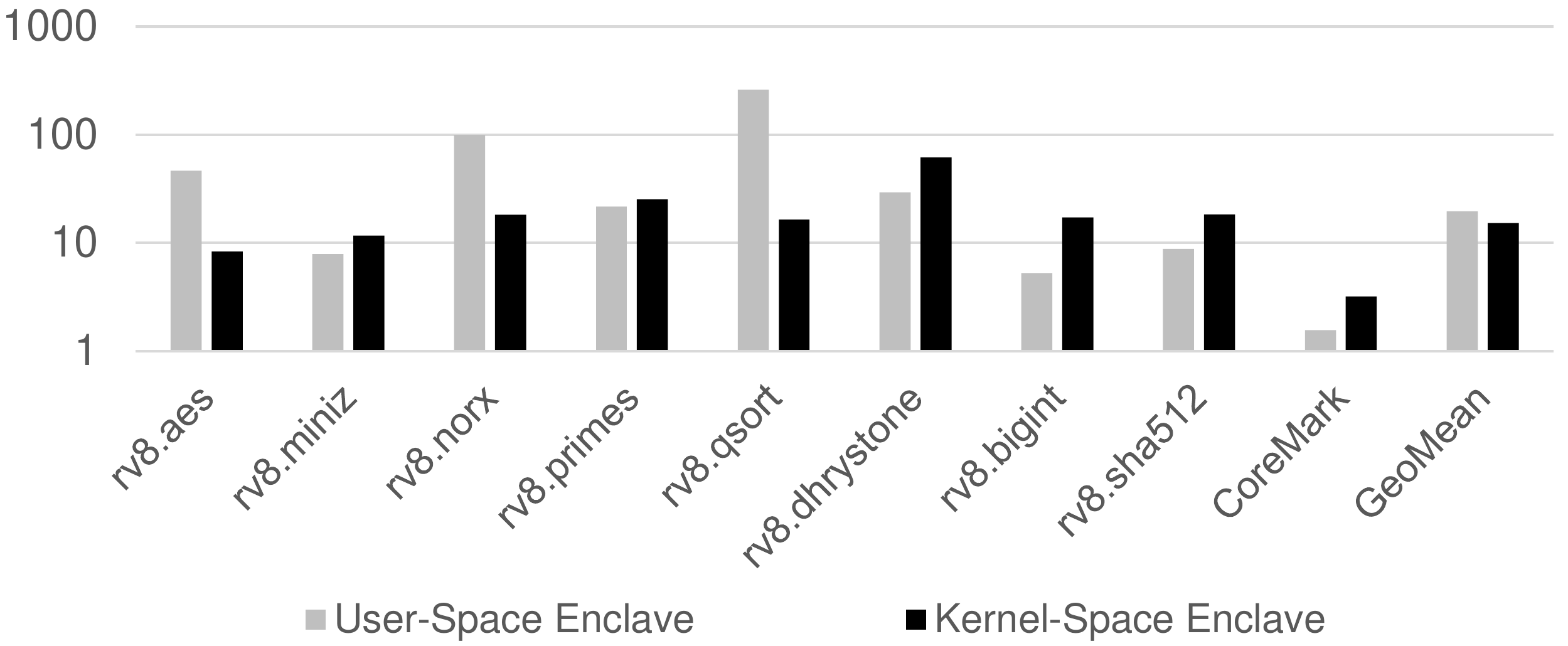}
	\caption{\arch performance overhead (in percent) on macro benchmarks \texttt{rv8} and \texttt{CoreMark} relative to a normal process.}
	\label{fig:macrorv8}
\end{figure}

For the enclave setup, our results show that most of the time (91.3\% for user-space, 52.1\% for kernel-space enclaves) is spent on binary verification. The \textit{Others} measurement contains all remaining steps of the setup process, e.g., loading of the enclave binary, enclave configuration, flushing of the TLB and L1 cache and jumping into the enclave. During our evaluation, we use 32KB 8-way set associative L1 data and instructions caches and a TLB with 32 entries. The setup of the kernel-space enclave is more complex and includes additional setup steps, namely, freeing the core from pending processes, detaching the core from the OS, and booting the RT. In the teardown phase, zeroing the memory produces 39.9\% of the overhead for the user-space and  45,7\% of the overhead for the kernel-space enclave). The cleaning is more time consuming for the kernel-space enclave because of the larger enclave memory region. The \textit{Others} measurement contains additional steps, e.g., exiting the enclave and flushing the TLB and L1 cache. In the kernel-space enclave case, the core must additionally be rebooted.

As the RT in our prototype does not support a suspension mode (keeping the enclave in memory), we emulate the \textit{context switch to the OS} by performing a teardown without zeroing memory, and the \textit{context switch from the OS} by performing a setup phase without verifying the enclave binary. Suspending the enclave and restoring it should be faster than a regular shutdown and boot, thus, this represents a worst-case approximation. The context switching measurements also contain the overhead for flushing the TLB and L1 cache, for which we measure 28 cycles and 3141 cycles, respectively.

As new entries to the page tables need to be verified by the SM, user-space enclaves have a higher overhead for dynamic memory allocation. In the kernel-space enclave case, all page tables are included in the enclave memory and thus, do not need to be verified. During communication, the OS can directly access a process's memory, whereas the user-space enclave needs to copy the data to be shared to the shared memory region. The kernel-space enclave additionally has to request the shared memory from the SM, and the OS needs to be notified by the SM using an inter-process interrupt.

\vspace{-0.5cm}
\subsubsection{Macrobenchmarks}
To evaluate the performance overhead in realistic scenarios, we used three different benchmarking suites that stress single cores, multi-core setups with two cores under test, and how the enclaves influence an OS under load. Furthermore, we measure the performance impact of our L2 cache partitioning by assigning $1/16$ of the L2 cache to the enclave under test.

\paragraph{Single-core benchmarks.} 
For single-core performance, we evaluated \arch with the RISC-V benchmark suites \texttt{rv8}~\cite{rv8} and \texttt{CoreMark}~\cite{coremark}, which are commonly used for TEE architectures~\cite{costan2016sanctum,keystone}. The results depicted in~\Cref{fig:macrorv8} are normalized to a normal user-space process. We measured a geometric mean of 19.70\% for user-space enclaves and 15.33\% for kernel-space enclaves for the performance overhead. As shown in~\Cref{tab:micro}, kernels-space enclaves have an increased setup time which however, amortizes with longer enclave run times. 
Outliers like \texttt{aes}, \texttt{norx} and \texttt{qsort} are memory-intensive workloads that perform a large number of context switches to the OS, mainly for dynamic memory allocation. Performing context switches and dynamic memory allocation is more expensive for the user-space enclave since the SM must verify newly created page table entries and copy them to the enclave memory. During one run, we count 24,601 syscalls for \texttt{aes}, 24,602 syscalls for \texttt{norx} and 48,846 syscalls for \texttt{qsort}. We also measure the overhead for flushing the TLB and L1 on every context switch which is, however, only necessary for the user-space enclave. The flushing induces only a small overhead which makes up for 1.03\%, 1.48\% and 1.21\% of the overall overhead for \texttt{aes}, \texttt{norx} and \texttt{qsort}, respectively.

\begin{table}[h!]
	\centering
	\small
	\renewcommand\theadgape{\Gape[0pt]}
	\setlength{\abovecaptionskip}{5pt plus 0pt minus 0pt}
	\begin{tabular}{c|rr}
		\textbf{Load/Cores}  & \multicolumn{1}{l}{\textbf{\thead{Normal\\Process}}} & \multicolumn{1}{l}{\textbf{\thead{Kernel-Space\\ Enclave}}}\\\hline
		30/1  & 1.49  & 1.49 (+-0.00\%) \\
		30/2  & 0.75  & 0.78 (+4.00\%)  \\
		500/1 & 27.65  & 28.82 (+4.23\%)  \\
		500/2 & 14.42 & 14.60 (+1.25\%)  \\
		1000/1 & 56.00 & 55.28 (-1.29\%) \\
		1000/2 & 27.64 & 27.81  (+0.62\%)  \\
		1500/1 & 83.62 & 83.64 (+0.02\%) \\
		1500/2 & 41.82 & 42.62 (+1.91\%)  \\
		2000/1 & 111.70 & 111.99 (+0.26\%)  \\
		2000/2 & 56.00 & 57.62 (+2.89\%)  \\
		\hline
		GeoMean &   -    &  +0.9\%       \\
		\hline
	\end{tabular}%
	\caption{Kernel-space enclave performance on multi-core \texttt{stress-ng} benchmark in seconds.}
	\label{tab:macro_stress}%
\end{table}%

\paragraph{Multi-core benchmarks.} 
Since \arch allows to assign multiple core to a kernel-space enclave, we evaluated \arch also on the dedicated multi-core benchmark \texttt{stress-ng}~\cite{stress-ng}. The results in \Cref{tab:macro_stress} show that multi-core kernel-space enclaves are practical by achieving almost the same performance as normal processes.

\paragraph{Influence on OS.} 
We stress the OS by running \texttt{CoreMark}, while starting an enclave in parallel. For the user-space enclave we use a single core, while two cores are needed for the kernel-space enclave, for which we simulate the suspension mode as in the microbenchmarks. {For one core, the CoreMark running on the OS is slowed down by 0.519s (1.56\%). For two cores with only one call after setting up the kernel-space enclave, the OS is slowed down by 0.792s (4.23\%), showing that the kernel-space enclave has a higher performance impact on the OS than the user-space enclave. Based on these results, we demonstrate that \arch also fulfills~\ref{req:frperfoverhead4} and achieves a moderate performance overhead.

\paragraph{L2 cache partitioning.} We evaluate the performance impact of partitioning the L2 cache (\texttt{CP-STRICT} mode) for kernel-space enclaves and show our results in~\Cref{tab:cache}. For our cycle-accurate experiments, we configure the core with 64KB 8-way set-associative L1 data and instructions caches and 2048KB 16-way set-associative shared L2 cache. The impact of way-based cache partitioning on performance is very application-dependent (besides the caches configuration and caches and main memory access latencies), as demonstrated by our experiments where the performance overhead ranges from a little under 0.2\%, as for the \texttt{prime} benchmark, to a little over 9\% for the \texttt{bigint} benchmark, for example. We measure a geometric mean of 3.09\%.
We note that the overheads reported are performance hits where the baseline is a best-case scenario where the only workload utilizing the cache resources (all 16 ways of the L2 cache) is the kernel-space enclave under test. Furthermore, we observe that performance significantly improves once more than 1 way is allocated per enclave, which is the likely scenario for enclaves that run applications with larger working sets and can benefit more from increased L2 cache resources.

\begin{table}[t]
	\footnotesize
	\setlength{\abovecaptionskip}{5pt plus 0pt minus 0pt}
	\setlength{\belowcaptionskip}{-16pt plus 0pt minus 0pt}
	\begin{center}
		\begin{tabular}{c|c|c|c}
			\multirow{2}{*}{\textbf{Benchmark}} & \textbf{Cycles \# for 16/16} & \textbf{Cycles \# for 1/16} & \textbf{Overhead}\\
			& \textbf{ways (baseline)} & \textbf{ways (worst-case)} & \textbf(+\%)\\
			\hline
			\hline
			rv8.aes & 29,754,631,670 &  32,175,733,155 & 8.1\% \\
			rv8.miniz & 42,040,536,353 & 45,063,752,315& 7.2\%\\
			rv8.norx &  30,899,386,564 & 32,702,249,193 & 5.8\%\\
			rv8.primes &  21,731,621,683 & 21,770,731,965 & 0.18\% \\
			rv8.qsort  & 24,355,792,115 & 25,280,228,818 & 3.8\% \\
			rv8.dhrystone & 19,865,586,529 & 20,289,555,571 & 2.1\% \\
			rv8.bigint & 65,512,466,917 & 71,487,944,568 & 9.1\%\\
			CoreMark & 394,664,199 & 402,293,814 & 	1.9\% \\
			\hline
			GeoMean & - & - & 3.09\% \\
		\end{tabular}
		\caption{Performance impact of L2 cache strict way-based partitioning for kernel-space enclaves on different benchmarks.}
		\label{tab:cache}
	\end{center}
\end{table}

\begin{table*}[ht]
	\centering
	\IfFileExists{stix2.sty}{%
		\newcommand{\half}{$\circlelefthalfblack$}
		\newcommand{\yes}{$\mdlgblkcircle$}
		\newcommand{\no}{$\mdlgwhtcircle$}%
	}{
		\newcommand{\half}{Incompl.}
		\newcommand{\yes}{Yes}
		\newcommand{\no}{No}%
		\textbf{You are missing the package "stix2-otf", thus, you are seeing no icons in the table.}
	}	
	\renewcommand\theadalign{bc}
	\renewcommand\theadfont{\bfseries}
	\renewcommand\theadgape{\Gape[1pt]}
	\renewcommand\cellgape{\Gape[0pt]}
	\newcommand{\invast}{\textcolor{white}{$^\ast$}} 
	\newcolumntype{C}{>{\footnotesize}c}
	\setlength{\abovecaptionskip}{2pt plus 0pt minus 0pt}
	\resizebox{\textwidth}{!}{
		\begin{tabular}{|c|C|c|c|c|c|c|c|}
			\hline
			&       &       \multicolumn{3}{c|}{\multirow{2}{*}{\thead{Enclave Type}}} &    &       &  \\
			\textbf{Name} & \multicolumn{1}{l|}{\thead{Extensions}} & \multicolumn{1}{c}{\thead{User-Space}} & \multicolumn{1}{c}{\thead{Kernel-Space}} & \multicolumn{1}{c|}{\thead{Sub-Space}} & \thead{Dynamic Cache\\Side-Channel Resilience} & \thead{Controlled Side-\\ Channel Resilience} & \thead{Enclave-to-Peripheral\\ Binding} \\
			\hline
			SGX~\cite{sgx1}   & \cite{ahmad2018obliviate,deja_vu,drsgx,varys,tsgx,ahmad2019obfuscuro} & \yes\invast   & \no\invast    & \no\invast  & \half$^\ast$   & \half$^\ast$   & \no\invast  \\
			Sanctum~\cite{costan2016sanctum} &   -   & \yes\invast   & \no\invast    & \no\invast   & \half\invast   & \yes\invast   &  \no\invast \\
			SEV(-ES)~\cite{amd_sev}   &   -    & \no\invast    & \yes\invast   & \no\invast    & \no\invast  & \no\invast    & \no\invast  \\
			TrustZone~\cite{arm-trustzone} & \cite{optee,sanctuary,komodo,hua2017vtz,trustice,guan2017trustshadow,tzasc_exynos5,zhao2019sectee} & \no\invast    & \yes\invast   & \no\invast   & \half$^\ast$   & \yes\invast    & \half\invast  \\
			Keystone~\cite{keystone} &   -    & \no\invast    & \yes\invast  & \no\invast    & \yes\invast   & \yes\invast   & \no\invast  \\
			\hline
			\large\arch &   -    & \yes\invast   & \yes\invast   & \yes\invast   & \yes\invast   & \yes\invast   & \yes\invast  \\
			\hline
		\end{tabular}%
	}
	\caption{Comparison of major TEE architectures with respect to provided enclave types, dyn. cache-side channel and controlled-side channel resilience, and enclave-to-peripheral binding, i.e., MMIO/DMA protection with exclusive enclave assignment. \yes~indicates full support, \half~for support with limitations, \no~for no support, $^\ast$~if resilience can only be achieved through extensions.}
	\label{tab:relatedwork}
\end{table*}%

%% file: sections/related_work.tex
\vspace{-0.4cm}
\section{Related Work}
\label{sec:related_work}
The existing works mostly related to \arch are TEE architectures which focus on modern high-performance computer systems. In contrast to capability systems or memory tagging extensions~\cite{woodruff2014cheri, vilanova2014codoms, erim, frassetto2018imix, song2016hdfi}, TEE architectures protect sensitive services in security contexts (enclaves) against privileged software adversaries. We do not further discuss TEE architectures focusing on embedded systems~\cite{noorman2013sancus,brasser2015tytan, koeberl2014trustlite, timberv}.

We compare \arch to other TEE architectures in~\Cref{tab:relatedwork}. All presented architectures provide a single type of enclave which, on an abstract level, resemble either the user-space or kernel-space enclaves provided by \arch.

Intel SGX~\cite{sgx1} offers user-space enclaves on Intel processors. The untrusted OS provides memory management and other OS services, e.g. exception handling, to the enclaves. SGX does not protect against cache side-channel~\cite{brasser2017software, cache_attack_btb} and controlled side-channel attacks~\cite{controlled_sc, van2017telling, van2018nemesis}. Many extensions to SGX were proposed in order to mitigate side-channel attacks~\cite{varys, tsgx, deja_vu, drsgx, ahmad2018obliviate, ahmad2019obfuscuro}, however, these solutions are all ad-hoc approaches that do not fix the underlying design shortcomings of SGX, but instead leverage costly data-oblivious algorithms~\cite{drsgx, ahmad2018obliviate, ahmad2019obfuscuro}, or exploit not commonly available hardware in an unintended way~\cite{tsgx, deja_vu}.

Sanctum~\cite{costan2016sanctum}, which also provides user-space enclaves, addresses both, cache side-channels through page coloring, and controlled side-channels by storing the enclave page tables in the enclave memory, like \arch. However, page coloring is not practical as it influences the whole OS memory layout and cannot be efficiently changed at run time. \arch's cache partitioning instead allows dynamic assignment of cache ways, and also mechanisms to mitigate interrupt-based side-channel attacks.
Sanctum and SGX only provide user-space enclaves which are inherently limited as they cannot provide secure I/O, but only protect from simple DMA attacks.

Similar to SGX, AMD SEV~\cite{amd_sev}, which isolates complete VMs in the form of kernel-space enclaves, does not consider any side-channel attacks. VM data in the CPU cache is protected by an access control mechanism relying on Address Space Identifiers which, however, does not protect against cache side-channel attacks. As the memory management and I/O services are provided by the untrusted hypervisor, SEV is also vulnerable to controlled side-channel attacks~\cite{severed_amd} and cannot provide secure peripheral binding~\cite{amd_io}. 

ARM TrustZone~\cite{arm-trustzone} separates the system into normal and secure world, a single kernel-space enclave which does not rely on the OS and thus, is protected from controlled side-channel attacks. TrustZone does not provide cache side-channels protection, only by using additional hardware~\cite{zhao2019sectee}. Further, TrustZone's major design shortcoming is providing only a single enclave, thus, sensitive services cannot be strongly isolated with TrustZone, hence, access to TrustZone is highly limited in practice by device vendors. Extensions building upon TrustZone mostly tried to enable multi-enclave support for TrustZone~\cite{sanctuary, hua2017vtz, trustice, tzasc_exynos5} with workarounds that either rely on ARM IP~\cite{sanctuary}, block the hypervisor~\cite{hua2017vtz,tzasc_exynos5}, or massively impact performance~\cite{trustice}. Since multiple enclaves were not considered in the TrustZone design from the beginning, even the proposed extensions cannot provide binding peripherals directly and exclusively to single enclaves.

Keystone~\cite{keystone} provides kernel-space enclaves on RISC-V. Moreover, Keystone uses a cache-way based partitioning against cache side-channel attacks, comparable to \arch. However, Keystone provides a coarse-grained cache ways assignment per CPU core, whereas \arch assigns cache ways to enclaves with freely configurable boundaries. Thus, the Keystone design is limited to a single enclave type which prevents Keystone from isolating the firmware from the actual TCB and demands adapting the sensitive services to the predefined enclave. Moreover, in contrast to \arch, Keystone does not support enclave-to-peripheral binding.

%% file: sections/conclusion.tex
\vspace{-0.5cm}
\section{Conclusion}
\label{sec:conclusion}
We presented \arch, a novel TEE architecture which provides strongly-isolated enclaves that can be adapted to the functionality and security requirements of the sensitive services which they protect. \arch offers different types of enclaves, ranging from sub-space enclaves, over user-space enclaves, to self-sustained kernel-space enclaves which can execute privileged software. \arch's protection mechanisms are based on new hardware security primitives on the system bus, the shared cache and the CPU. We instantiate \arch on a RISC-V system. The evaluation of our prototype indicates minimal hardware overhead for the security primitives and a moderate overall performance overhead.

\vspace{-0.3cm}
\section*{Acknowledgments}
We thank our anonymous reviewers for their valuable and constructive feedback.
This work was funded by the Deutsche Forschungsgemeinschaft (DFG) – SFB 1119 – 236615297. Moreover, this project has received funding from Huawei within the OpenS3 lab.